\documentclass[aps, prb, twocolumn, amssymb, amsmath, showpacs, superscriptaddress]{revtex4-1}

\usepackage{bm}
\usepackage{graphicx}
\usepackage{dcolumn}
\usepackage{graphicx}
\usepackage[colorlinks,citecolor=blue,linkcolor=blue]{hyperref}

\begin{document}

\title{Intrinsic spin Hall effect in monolayers of group-VI dichalcogenides: A first-principles study}

\author{Wanxiang Feng}
\affiliation {School of Physics, Beijing Institute of Technology, Beijing 100081, China}
\affiliation {Materials Science \& Technology Division, Oak Ridge National Laboratory, Oak Ridge, TN 37831, USA}
\affiliation {Department of Physics and Astronomy, University of Tennessee, Knoxville, TN 37996, USA}

\author{Yugui Yao}
\email{ygyao@bit.edu.cn}
\affiliation {School of Physics, Beijing Institute of Technology, Beijing 100081, China}

\author{Wenguang Zhu}
\affiliation {Department of Physics and Astronomy, University of Tennessee, Knoxville, TN 37996, USA}
\affiliation {Materials Science \& Technology Division, Oak Ridge National Laboratory, Oak Ridge, TN 37831, USA}

\author{Jinjian Zhou}
\affiliation {Institute of Physics, Chinese Academy of Sciences and Beijing National Laboratory for Condensed Matter Physics, Beijing 100190, China}
\affiliation {School of Physics, Beijing Institute of Technology, Beijing 100081, China}

\author{Wang Yao}
\affiliation {Department of Physics and Center of Theoretical and Computational Physics, The University of Hong Kong, Hong Kong, China}

\author{Di Xiao}
\email{dixiao@cmu.edu}
\affiliation {Materials Science \& Technology Division, Oak Ridge National Laboratory, Oak Ridge, TN 37831, USA}
\affiliation {Department of Physics, Carnegie Mellon University, Pittsburgh, PA 15213, USA}

\date{\today}

\begin{abstract}
Using first-principles calculations within density functional theory,
we investigate the intrinsic spin Hall effect in monolayers of group-VI transition-metal dichalcogenides $MX_2$ ($M = \text{Mo, W}$ and  $X = \text{S, Se}$).  $MX_2$ monolayers are direct band-gap semiconductors with two degenerate valleys located at the corners of the hexagonal Brillouin zone.  Because of the inversion symmetry breaking and the strong spin-orbit coupling, charge carriers in opposite valleys carry opposite Berry curvature and spin moment, giving rise to both a valley-Hall and a spin-Hall effect.  We also show that the intrinsic spin Hall conductivity in inversion-symmetric bulk dichalcogenides is an order of magnitude smaller compared to monolayers.  Our result demonstrates monolayer dichalcogenides as an ideal platform for the integration of valleytronics and spintronics.
\end{abstract}


\maketitle

\section{INTRODUCTION}

In crystalline semiconductors, it often happens that the conduction band minima and valence band maxima are located at degenerate but inequivalent valleys.  Well-known examples include graphene,~\cite{Rycerz2007} bismuth thin films,~\cite{Zhu2012} and AlAs quantum wells.~\cite{Gunawan2006}  Since the valleys are usually separated by a large distance in momentum space, intervalley scattering is greatly suppressed in the presence of smooth scattering potential, rendering the valley index an intrinsic property for low-energy carriers. Motivated by this observation, there has been a growing interest in exploiting the valley index in electronic devices, much in the same way as the spin index is used in spintronic applications. This is the subject of valleytronics.

Recently, a general scheme based on inversion symmetry breaking has been proposed to generate and manipulate the valley polarization.~\cite{Xiao2007,Yao2008}   The central idea is that under inversion symmetry breaking, the valley index can be associated with distinctive physical quantities such as the Berry curvature and orbital magnetic moment.~\cite{Xiao2010}  Using graphene as an example, the authors of Ref.~\onlinecite{Xiao2007,Yao2008} showed that inversion symmetry breaking allows a valley Hall effect in which carriers in different valleys flow to opposite transverse edges when an electric field is applied, leading to a finite valley polarization along the edges.~\cite{Xiao2007} Furthermore, it also gives rise to valley-contrasting circular dichroism in the momentum space, which takes the extreme form of optical selection rules at high symmetry points.~\cite{Yao2008} Other approaches have also been proposed.~\cite{Rycerz2007,Gunlycke2011}  However, they all rely on carefully prepared geometry at the atomic scale, which is difficult to control in experiments.

In general, inversion symmetry breaking also lifts the spin degeneracy of energy bands in the presence of spin-orbit coupling (SOC). As required by time-reversal symmetry, the spin-splitting in opposite valleys must be opposite, therefore the valley carriers can be also distinguished by their spin moments. This is the basis of \emph{coupled} spin and valley physics. However, the SOC is negligibly small in graphene,~\cite{Min2006,Yao2007} preventing further investigation along this direction. In a recent work,~\cite{Xiao2012} we have studied monolayers of group-VI transition-metal dichalcogenides for the following reasons: (i) the inversion symmetry is explicitly broken in monolayers; (ii) the conduction and valence bands of these materials harbors a multi-valleyed structure;~\cite{Li2007,Lebegue2009} and (iii) the SOC is substantial due to the presence of heavy metal atoms.~\cite{Zhu2011}  Therefore these materials provide a perfect platform to investigate the interplay between spin and valley degrees of freedom. Based on an effective $k\cdot p$ model, we predicted that the valley Hall effect is accompanied by a spin Hall effect in both $n$- and $p$-doped systems, and the valley-dependent optical selection rule also becomes spin-dependent.~\cite{Xiao2012}

Monolayers of group-VI dichalcogenides also display excellent optical properties for practical applications.  Recent experiments have demonstrated that MoS$_2$, a prototypical group-VI dichalcogenide, crossovers from an indirectgap semiconductor at multilayers to a direct band-gap one at monolayer.~\cite{Splendiani2010,Mak2010}  The direct band-gap is in the visible frequency range, most favorable for optoelectronic applications. Experimental evidence of the valley-dependent optical selection rule in monolayer MoS$_2$ has been recently reported based on polarization-sensitive photoluminescence measurement.~\cite{Zeng2012,Mak2012,Cao2012}

In this work we present a comprehensive first-principles study of the coupled spin and valley physics, focusing on the Hall effects of valley and spin.  We show that, because of the inversion symmetry breaking and the strong SOC, charge carriers in opposite valleys carry opposite Berry curvature and spin moment, giving rise to both the valley- and spin-Hall effect.  Our first-principles calculations provide a quantitative basis for the $k\cdot p$ model derived in Ref.~\onlinecite{Xiao2012}, and subtle differences between these two are discussed.  We also show that the intrinsic spin Hall conductivity in inversion-symmetric bulk dichalcogenides is an order of magnitude smaller compared to monolayers.  Our result demonstrates monolayer dichalcogenides as an ideal platform for the integration of valleytronics and spintronics.

\section{Methodology}

The electronic ground-state calculations in this work were performed using full-potential linearized augmented plane-wave method,~\cite{Singh1994} implemented in the package \textsc{wien2k}.~\cite{Blaha2001}  Exchange-correlation effect was treated with the Perdew, Burke, and Ernzerhof parameterized generalized-gradient approximation.~\cite{Perdew1996}  The crystal structure were adopted from the first-principles optimized results.~\cite{Zhu2011}  For the slab model, a 20 \AA \ thick vacuum layer was used to avoid the interactions between adjacent monolayers.  The converged ground-states were obtained using $k$-mesh $16\times16\times3$ for bulk and $16\times16\times1$ for monolayer in the first Brillouin zone, both with $K_\text{max}R_\text{MT}=7.0$, where $R_\text{MT}$ represents the smallest muffin-tin radius and $K_\text{max}$ is the maximum size of reciprocal-lattice vectors.  Wave functions and potentials inside the atomic sphere were expanded in spherical harmonics up to $l$=10 and 4, respectively.  Spin-orbit coupling was included by a second-variational procedure,~\cite{Singh1994} where states up to 9 Ry above Fermi level were included in the basis expansion.

To calculate the Berry curvature and Hall conductivity, we first computed the Wannier functions by  the maximally localized algorithm,~\cite{Marzari1997,Souza2001} implemented in the package \textsc{wannier90}.~\cite{Mostofi2008}  The construction of maximally localized Wannier functions is a non-self-consistent process on a uniform $8\times8\times8$ grid of $k$-point with formerly converged self-consistent charge potential. Transition-metal dichalcogenides have the chemical formula $MX_2$ ($M = \text{Mo, W}$ and  $X = \text{S, Se}$).  In the case of monolayers, there are 22 bands in the energy range from about -6 to 5 eV, mainly formed by $M$ $d$- and $X$ $p$-orbitals.  With this in mind, we chose ten $d$ orbitals on atom $M$ and six $p$ orbitals on each atom $X$ as the initial guess of the Wannier functions.  After less than 200 iterative steps, the total Wannier spread was well converged down to 10$^{-7}$ Bohr$^2$.  On the other hand, in bulk systems, the unit cell contains two formula units, and there are 44 bands also formed by $M$ $d$- and $X$ $p$-orbitals.  The construction process in bulk is similar to that in monolayers except for WS$_2$ and WSe$_2$.  In these materials, the conduction bands are entangled with higher bands.  The disentanglement approach~\cite{Souza2001} was applied to bulk WS$_2$ and WSe$_2$.  Once the Wannier functions were obtained, we followed Ref.~\onlinecite{Wang2006} to calculate the Berry curvature and integrate it over the Brillouin zone to obtain the Hall conductivity.

\begin{figure}
\includegraphics[width=\columnwidth]{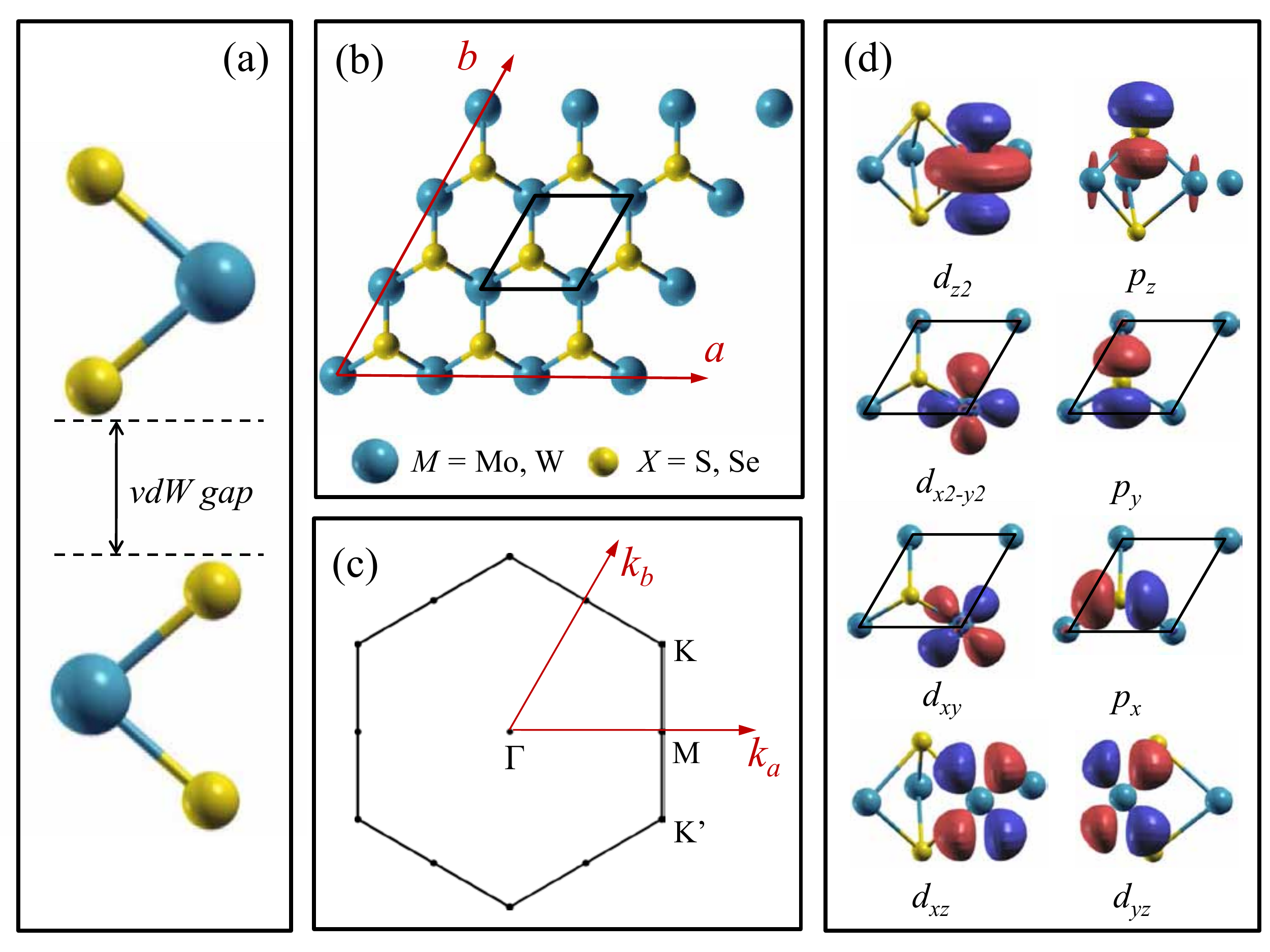}
\caption{(Color online)  (a) Side view of the unit cell of the 2$H$-$MX_2$ structure with $M$=Mo, W and $X$=S, Se.  It contains two $MX_2$ monolayers separated by a Van der Waals gap. (b) Top view of the $MX_2$ monolayer.  The black lines indicate the unit cell in $ab$ plane.  (c) The first Brillouin zone and high symmetry points of the $MX_2$ monolayer.  (d) The Wannier functions of the $MX_2$ monolayer, including five $d$-orbitals on $M$ atom and three $p$-orbitals on each $X$ atom.}
\label{fig:crystal}
\end{figure}

\section{Band Structure and Berry Curvature}

In this section we present the electronic band structure and Berry curvature of monolayers of group-VI dichalcogenides, using MoS$_2$ as an example.  We show that inversion symmetry breaking gives rise to two physical quantities, the spin moment and the Berry curvature, that can be used to distinguish valley carriers.

Structurally, MoS$_2$ can be regarded as strongly bonded two-dimensional S-Mo-S layers that are loosely coupled to one another by Van der Waals interactions.  Within each layer, the Mo and S atoms form hexagonal lattices in separate planes with each Mo atom coordinated by six nearest-neighboring S atoms in the trigonal prismatic geometry [Fig.~\ref{fig:crystal}(a) and (b)].  In its bulk form, MoS$_2$ has the so-called $2H$ stacking order with space group $P6_3/mmc$ ($D_{6h}^4$), which is inversion symmetric [Fig.~\ref{fig:crystal}(a)]. Because of the weak interlayer coupling, this layered compound can be easily exfoliated into monolayers by mechanical~\cite{Novoselov2005,Radisavljevic2011} and chemical~\cite{Coleman2011} means.  In monolayer MoS$_2$, the space group is reduced to $P\bar{6}m2$ ($D_{3h}^1$) with explicit breaking of inversion symmetry.

\begin{figure}
\includegraphics[width=\columnwidth]{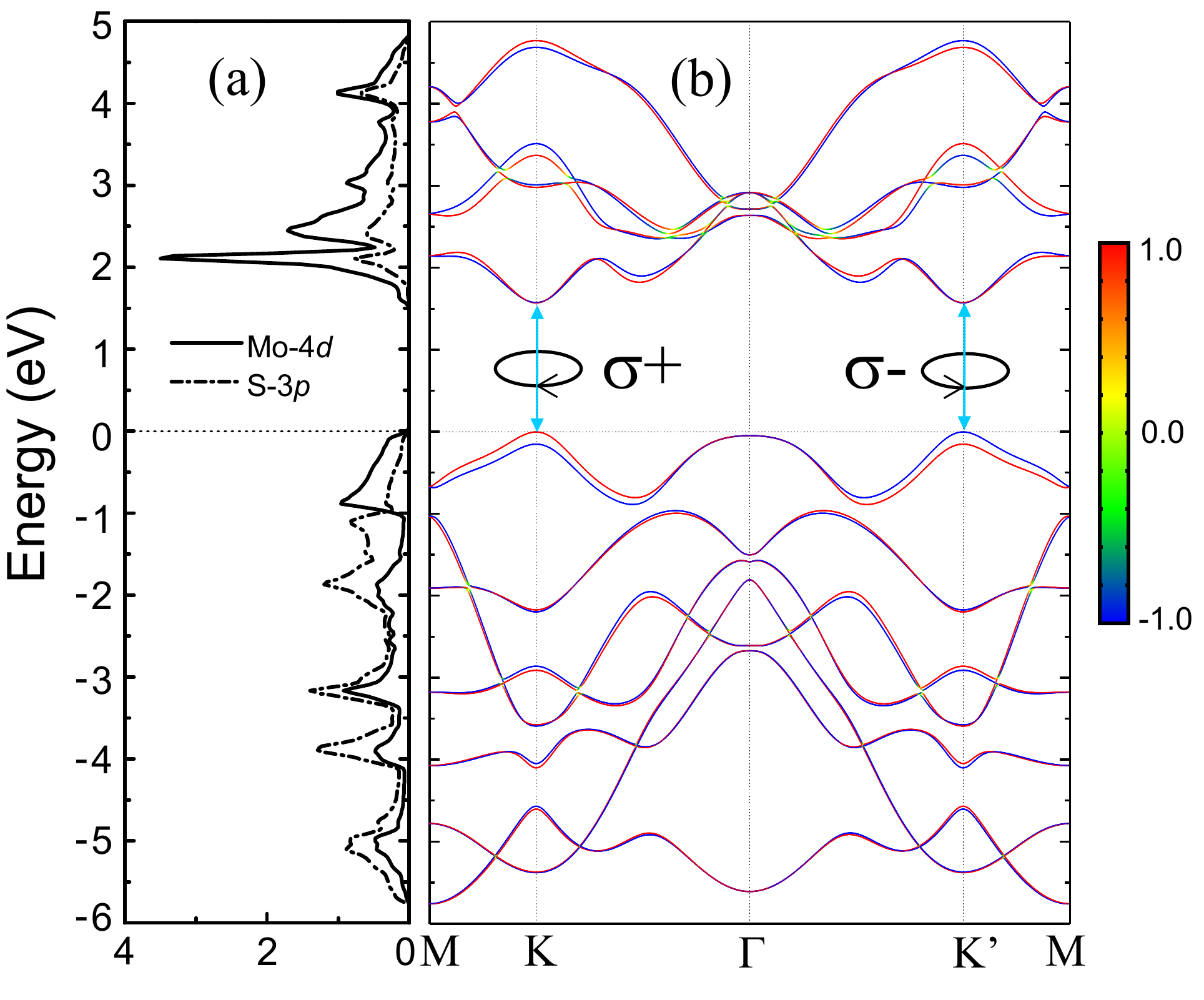}
\caption{(Color online)  The calculated electronic structure of MoS$_2$ with the spin-orbit coupling.  (a) The partial density of states for Mo-4$d$ and S-3$p$ orbitals, respectively, in the unit of states/eV/cell.  (b) The band structure with the projection of spin operator $\hat{s}_{z}$ (color map).  The red and blue colors indicate the spin-up and -down states, respectively.  The optical transitions between the VBM and the CBM are coupled exclusively with $\sigma+$ ($\sigma-$) circular polarizations at the inequivalent valleys $K$ ($K^{\prime}$).~\cite{Xiao2012} }
\label{fig:band}
\end{figure}

Before moving on to the discussion of coupled spin and valley physics, we briefly study the orbital characters via the Wannier functions.  The partial density of states of monolayer MoS$_2$ shown in Fig.~\ref{fig:band}(a) clearly shows that in the energy range of -6 to 5 eV the contributions to electronic states mainly come from the Mo $d$- and S $p$-orbitals, whereas other orbitals have vanishing contributions.  In Fig.~\ref{fig:crystal}(d), we plot the typical Wannier functions for five Mo $d$- and three S $p$-orbitals together with the unit cell in real space.  If the SOC is turned on, the number of orbitals will be doubled and altogether 22 bands are formed.  Under the crystal field of trigonal prismatic coordination, the $d$-orbitals split into $\{d_{z^2}\}$, $\{d_{xz}, d_{yz}\}$, $\{d_{x^2-y^2}, d_{xy}\}$ and $p$-orbitals split into $\{p_{z}\}$, $\{p_{x}, p_{y}\}$.  At the two inequivalent valleys, $K$ and $K'$, the valence band maximum (VBM) is constructed by the Mo $\{d_{x^2-y^2}, d_{xy}\}$ orbitals with some mixing from the S $\{p_{x}, p_{y}\}$ orbitals, while the conduction band minimum (CBM) is dominated by Mo $d_{z^2}$ orbitals. These orbital characters of band-edges are in line with the analysis of the $k\cdot p$ model in Ref.~\onlinecite{Xiao2012}.

Figure~\ref{fig:band}(b) shows the fully relativistic band structure of monolayer MoS$_2$ with the projection of spin operator $\hat{s}_{z}$, i.e., $\left\langle \psi_{n\bm{k}}\right|\hat{s}_{z}\left|\psi_{n\bm{k}}\right\rangle $, obtained from Wannier interpolation.~\cite{Yates2007}  We can see that there is a direct band-gap at the two inequivalent corners $K$ and $K'$ of the Brillouin zone.  Furthermore, a large spin splitting ($\sim$ 150 meV) appears at the VBM with opposite spin moments at the two valleys, as a result of inversion symmetry breaking.~\cite{Zhu2011}  This indicates that in addition to their valley index, the valley carriers in the valence bands can be also distinguished by their spin index.  On the other hand, since the CBM state is made of the Mo $d_{z^2}$ orbital, SOC is inactive and the CBM remains degenerate (to first order of the SOC).

\begin{figure}
\includegraphics[width=\columnwidth]{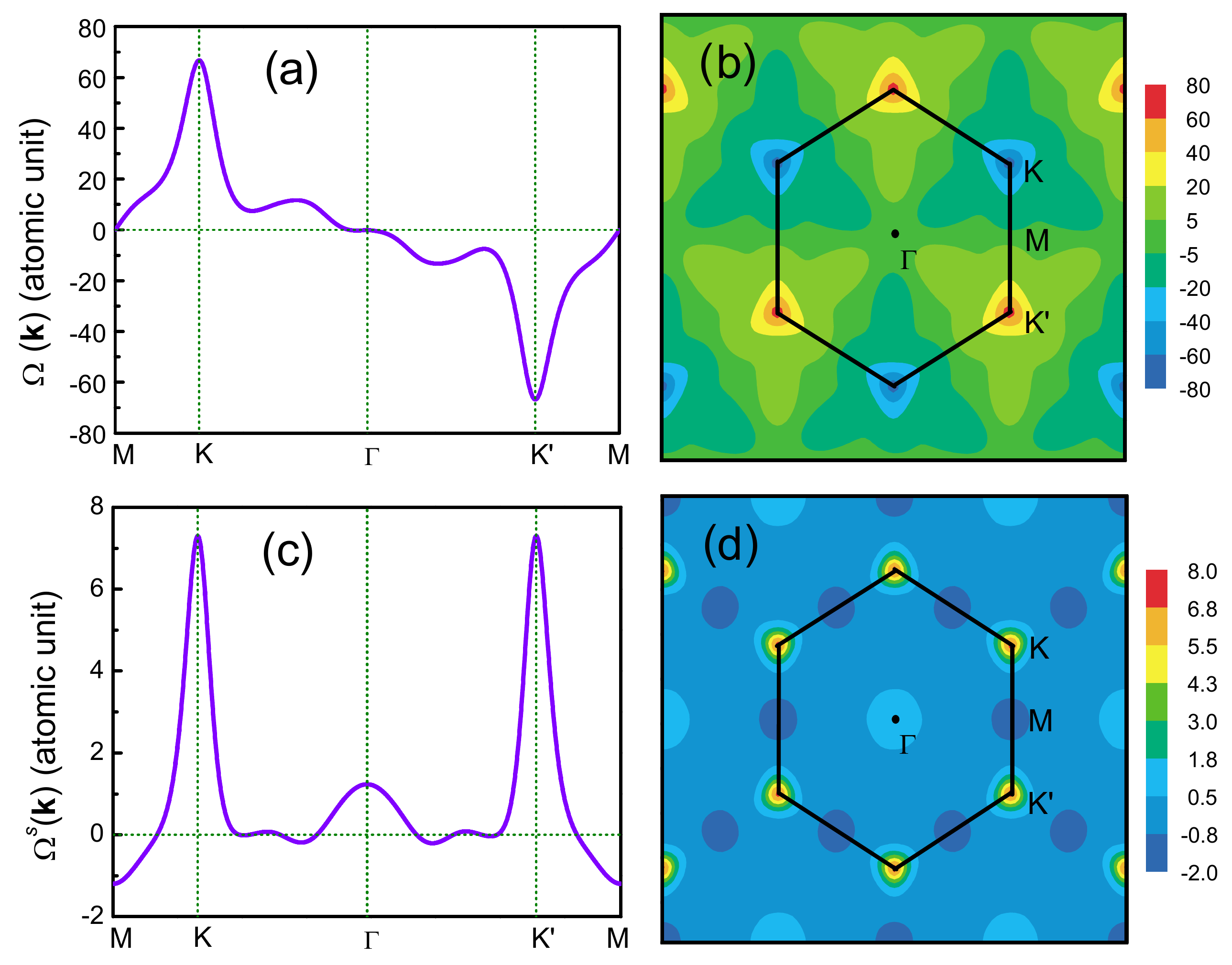}
\caption{(Color online)  The Berry curvatures of monolayer MoS$_2$ along the high symmetry lines (a) and in the 2D $k$-plane (b).  The spin Berry curvatures of monolayer MoS$_2$ along the high symmetry lines (c) and in the 2D $k$-plane (d).  All of the Berry curvatures are in the atomic unit (Bohr$^2$).}
\label{fig:berry}
\end{figure}

In the presence of inversion symmetry breaking, the charge carriers also acquire a valley-contrasting Berry curvature.~\cite{Xiao2007,Xiao2010}  According to the Kubo-formula,~\cite{Thouless1982,Yao2004} the Berry curvature $\Omega\left(\bm{k}\right)$ of the occupied states can be written as
\begin{equation} \label{eq:omega_k}
\Omega\left(\bm{k}\right)=\sum_{n}f_{n}\Omega_{n}\left(\bm{k}\right),
\end{equation} and
\begin{equation} \label{eq:omega_nk}
\Omega_{n}\left(\bm{k}\right)=-\sum_{n^{\prime}\neq n}\frac{2\textrm{Im}\left\langle \psi_{n\bm{k}}\right|\bm{v}_{x}\left|\psi_{n^{\prime}\bm{k}}\right\rangle \left\langle \psi_{n^{\prime}\bm{k}}\right|\bm{v}_{y}\left|\psi_{n\bm{k}}\right\rangle }{\left(E_{n^{\prime}}-E_{n}\right)^{2}},
\end{equation}
where $\left|\psi_{n\bm{k}}\right\rangle $ is the Bloch function with the eigenvalue $E_{n}$, $f_{n}$ the Fermi-Dirac distribution function, and $\bm{v}_{x(y)}$ the velocity operators.  Here, we have used the maximally localized Wannier functions as the basis to calculate the Berry curvature [Eq.~(\ref{eq:omega_nk})] and the spin Berry curvature [below in Eq.~(\ref{eq:omega_snk})].~\cite{Wang2006}  Figure~\ref{fig:berry}(a) shows $\Omega(\bm k)$ of monolayer MoS$_2$ along the high-symmetry lines.  We can see that $\Omega(\bm k)$ is significantly peaked at both $K$ and $K^{\prime}$ but with opposite signs.  The $k$-space contrasting $\Omega(\bm k)$ in systems without inversion symmetry is a key quantity to characterize the chirality of the Bloch electrons and is the basis for valley-contrasting phenomena.~\cite{Xiao2007,Yao2008,Xiao2012}  Away from the two valleys, $\Omega(\bm k)$ decays rapidly and vanishes at the $\Gamma$ and $M$ points.  We also plot the map distribution of $\Omega(\bm k)$ in the 2D $k$-plane, as shown in Fig.~\ref{fig:berry}(b), which clearly shows the $C_{3}$ symmetry of the system.  This is in contrast with an energy counter plot, which would display the $C_6$ symmetry, i.e., the two valleys are energetically indistinguishable, but they can be distinguished by their Berry curvatures.

\begin{figure}
\includegraphics[width=\columnwidth]{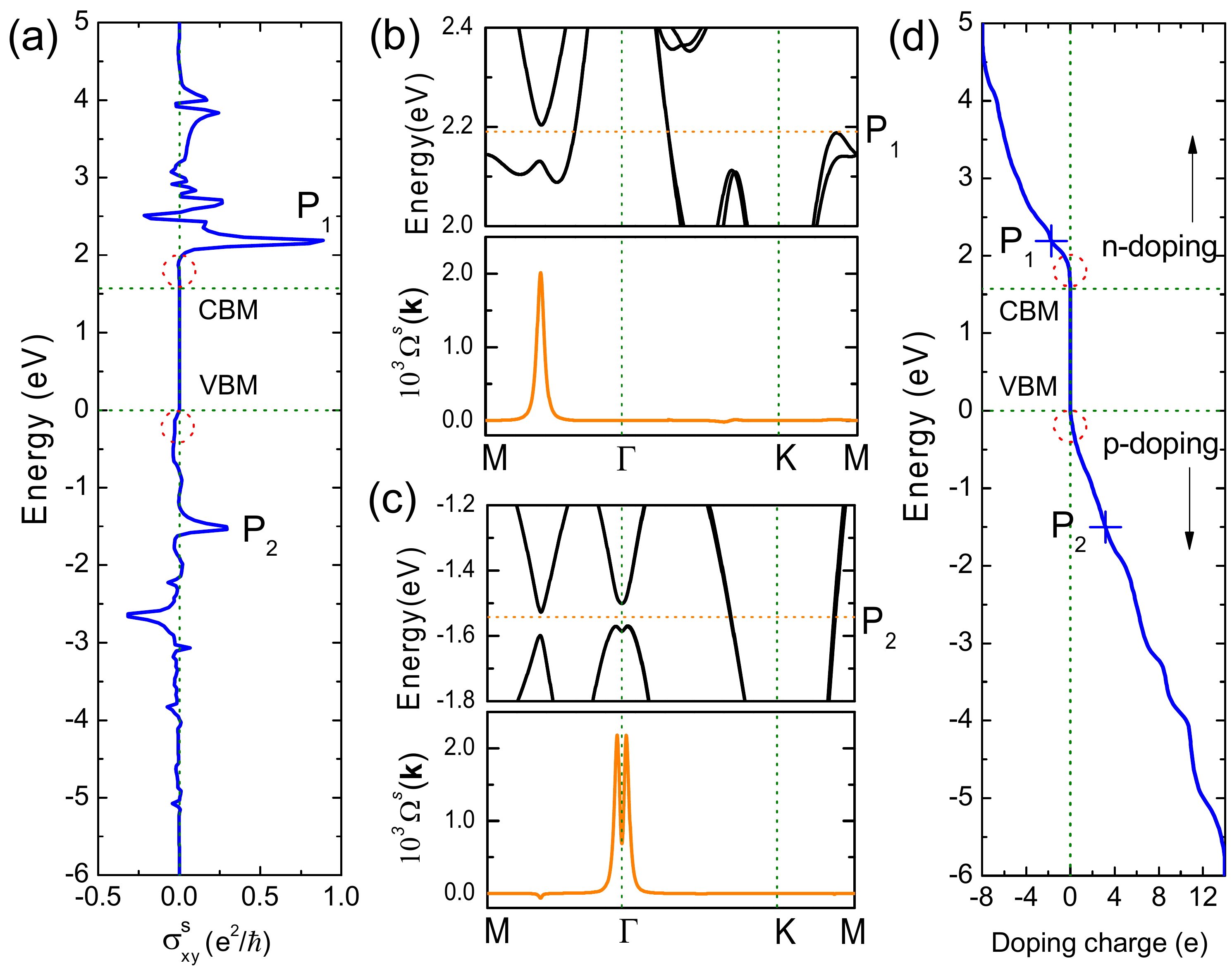}
\caption{(Color online)  (a) The intrinsic spin Hall conductivity $\sigma^{s}_{xy}$ ($e^2/\hbar$) as a function of the Fermi energy for monolayer MoS$_2$.  The energy zero point (true Fermi level) is at the VBM.  Two large peaks close to the CBM and VBM are denoted by $P_1$ and $P_2$, respectively.   (b)(c) The band structure (up panel) and the spin Berry curvature (down panel) when the Fermi level shifts to the positions of the peak $P_1$ and $P_2$, respectively.  There is a small peak of $\Omega^{s}$ with negative value along the $M$-$\Gamma$ line in (c). (d) The $n$- and $p$-doping charge as a function of the Fermi energy.  The low doping regimes just above the CBM and below the VBM (indicated by red circles) are more relevant in experiments.}
\label{fig:shc}
\end{figure}

\begin{table}[b]
\caption{\label{berry}Comparison of the Berry curvatures at VBM and CBM between the present first-principles calculations (the first line) and the $k\cdot p$ model in Ref.~\onlinecite{Xiao2012} (the second line).  $\Omega_{v(c)\uparrow(\downarrow)}$ is the Berry curvature of the valence (conduction) band with spin $\uparrow$ ($\downarrow$), given in the unit of Bohr$^2$.}
\begin{ruledtabular}
\begin{tabular}{cdddd}
\multicolumn{1}{r}{\textrm{}} &
\multicolumn{1}{r}{\textrm{MoS$_2$}} &
\multicolumn{1}{r}{\textrm{MoSe$_2$}} &
\multicolumn{1}{r}{\textrm{WS$_2$}} &
\multicolumn{1}{r}{\textrm{WSe$_2$}} \\
\hline
$\Omega_{v\uparrow}$     &  38.8 &  39.7 &  59.8 &  64.3 \\
                         &  35.3 &  36.5 &  55.4 &  60.0 \\
$\Omega_{v\downarrow}$   &  31.6 &  30.0 &  34.9 &  34.7 \\
                         &  29.5 &  28.4 &  34.2 &  33.3 \\
$\Omega_{c\uparrow}$     & -35.7 & -36.8 & -54.7 & -59.2 \\
                         & -35.3 & -36.5 & -55.4 & -60.0 \\
$\Omega_{c\downarrow}$   & -28.8 & -27.3 & -31.0 & -30.8 \\
                         & -29.5 & -28.4 & -34.2 & -33.3 \\
\end{tabular}
\end{ruledtabular}
\end{table}

The Berry curvature drives an anomalous transverse velocity in the presence of an electric field $\bm E$:~\cite{Xiao2010}
\begin{equation}
\bm v_\perp = -\frac{e}{\hbar} \bm E \times \bm\Omega(\bm k) \;,
\end{equation}
which is responsible for the intrinsic contribution to the anomalous Hall effect.~\cite{Yao2004}  However, in our systems, the charge carriers in the two valleys have opposite transverse velocities due to the opposite signs of the Berry curvatures.  Hence, the total anomalous Hall conductivity vanishes because of time reversal symmetry.  If a finite valley polarization can be generated, for example, by shining the sample with circularly polarized light, then a charge Hall current will appear.~\cite{Xiao2012}  On the other hand, since both the valley and spin current remains invariant under time-reversal, the valley Hall and the spin Hall effect can appear in time-reversal invariant systems, as long as the inversion symmetry is broken.~\cite{foot}

Finally, we compare the value of the Berry curvature at the VBM and CBM from both the effective $k\cdot p$ model~\cite{Xiao2012} and first-principles calculation in Table~\ref{berry}.  The excellent agreement between them further confirms the validity of the $k\cdot p$ model.

\section{The Intrinsic Spin Hall Effect}

As discussed above, both the valley Hall and spin Hall effect exist in $MX_2$ monolayers due to the valley-contrasting Berry curvature.  Note that the valley index is defined only in the vicinity of the valleys, whereas the spin index is defined everywhere in the Brillouin zone.  Therefore we will only calculate the intrinsic spin Hall conductivity.  For hole-doped samples, when the Fermi energy lies between the spin-split VBM states, the valley Hall conductivity coincides with the spin Hall conductivity.~\cite{Xiao2012}

At zero-temperature and clean limit, the intrinsic spin Hall conductivity (ISHC) tensor is given by
\begin{equation} \label{eq:sigma}
\sigma^{s}_{xy}=\frac{e}{\hbar}\int_{V_{G}}\frac{d^{2}k}{\left(2\pi\right)^{2}}\Omega^{s}\left(\bm{k}\right).
\end{equation}  For the convenience of discussion, in the following, we multiply a factor $2e/\hbar$ to the calculated ISHC to convert its unit to charge conductivity.  We can carry out the calculation of $\sigma^{s}_{xy}$ again using the Kubo-formulas~\cite{Yao2005,Guo2005}
\begin{equation} \label{eq:omega_sk}
\Omega^{s}\left(\bm{k}\right)=\sum_{n}f_{n}\Omega_{n}^{s}\left(\bm{k}\right),
\end{equation} and
\begin{equation}\label{eq:omega_snk}
\Omega_{n}^{s}\left(\bm{k}\right)=-\sum_{n^{\prime}\neq n}\frac{2\textrm{Im}\left\langle \psi_{n\bm{k}}\right|\bm{j}_{x}\left|\psi_{n^{\prime}\bm{k}}\right\rangle \left\langle \psi_{n^{\prime}\bm{k}}\right|\bm{v}_{y}\left|\psi_{n\bm{k}}\right\rangle }{\left(E_{n^{\prime}}-E_{n}\right)^{2}},
\end{equation} where $\bm{j}_{x}$ is the spin current operator defined as $\frac{1}{2}(\hat{s}_{z}v_{x}+v_{x}\hat{s}_{z})$.  We add a superscript $s$ for the spin Berry curvature in order to distinguish them from the ordinary Berry curvature in Eq.~(\ref{eq:omega_k}) and (\ref{eq:omega_nk}).  We can see that $\Omega^s(\bm k)$ of monolayer MoS$_2$ is peaked at both $K$ and $K^{\prime}$ with the same sign, as shown in Fig.~\ref{fig:berry}(c).  This can be understood as the following.  At the VBM, $s_z$ remains a good quantum number, and the spin Berry curvature is simply given by $\Omega^s(\bm K) = s_z\Omega(\bm K)$.  As both $s_z$ and $\Omega(\bm K)$ flip sign when $\bm K \to -\bm K$, $\Omega^s(\bm K)$ remains the same.  Figure~\ref{fig:berry}(d) shows the map distribution of $\Omega^s(\bm k)$ in the 2D $\bm k$-plane.  We observe that it has a clear $C_{6}$ symmetry rather than the $C_{3}$ symmetry in Fig.~\ref{fig:berry}(b).

\begin{figure}
\includegraphics[width=\columnwidth]{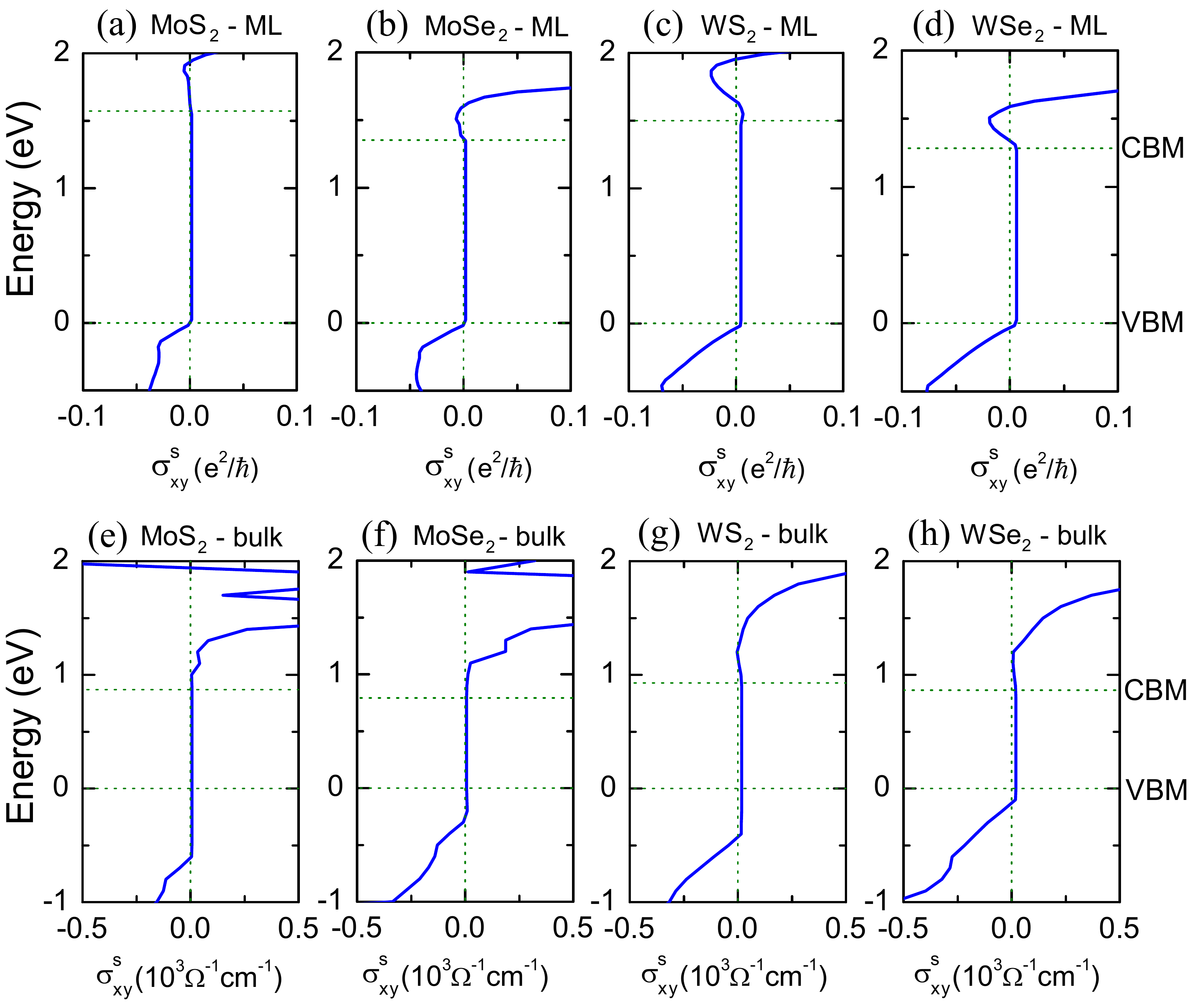}
\caption{(Color online)  The intrinsic spin Hall conductivity $\sigma^{s}_{xy}$ in the low doping regimes for the monolayer (a)-(d) and bulk (e)-(h) of the MoS$_2$, MoSe$_2$, WS$_2$, and WSe$_2$, respectively.  Dashed lines indicate the band-edges.  Note that the band-gaps of bulk are smaller than the monolayer ones.  The unit of $\sigma^{s}_{xy}$ is $e^2/\hbar$ ($\cong$ 2.43$\times$10$^{-4}\Omega^{-1}$) for 2D monolayer system, whereas $\Omega^{-1}$cm$^{-1}$ for 3D bulk system.  For quantitatively comparing the $\sigma^{s}_{xy}$ in bulk and monolayer, one needs to divide the $\sigma^{s}_{xy}$ in monolayer by its thickness.}
\label{fig:shc-2}
\end{figure}

\begin{table}[b]
\caption{The ISHC $\sigma^{s}_{xy}$ of monolayer $MX_2$ calculated at hole-doping concentration $n_{h}$=1.0$\times$10$^{13}$ cm$^{-2}$.  The slopes of $\sigma^{s}_{xy}$ when Fermi level lies inside the spin splitting gaps below the VBM are also listed.  The first and second lines are obtained from first-principles calculation and two-bands $k\cdot p$ model~\cite{Xiao2012}, respectively.}\label{tab:slope}
\begin{ruledtabular}
\begin{tabular}{cdddd}
\multicolumn{1}{r}{\textrm{}} &
\multicolumn{1}{r}{\textrm{MoS$_2$}} &
\multicolumn{1}{r}{\textrm{MoSe$_2$}} &
\multicolumn{1}{r}{\textrm{WS$_2$}} &
\multicolumn{1}{r}{\textrm{WSe$_2$}} \\
\hline
slope ($e^2/\hbar/$eV)    &  0.21 &  0.24 &  0.17  &  0.18   \\
                          &  0.20 &  0.23 &  0.20  &  0.23   \\
$\sigma^{s}_{xy}$ (10$^{-2}$ $e^2/\hbar$)   & -0.58  & -0.92  & -1.13  & -1.30  \\
                                               & -0.57  & -1.00  & -1.33  & -1.52  \\
\end{tabular}
\end{ruledtabular}
\end{table}

By integrating $\Omega^s(\bm k)$ over the occupied states, we obtain $\sigma^{s}_{xy}$ as a function of Fermi level for monolayer MoS$_2$, shown in Fig.~\ref{fig:shc}(a). Here, we set the energy zero point (true Fermi level) at the VBM, and calculate $\sigma^{s}_{xy}$ by rigidly shifting the Fermi level position.  For $n$-doped monolayer MoS$_2$, the calculated $\sigma^{s}_{xy}$ sharply reaches its maximum value of 0.89 $e^2/\hbar$ at 2.19 eV.  When further increasing the doping concentration, $\sigma^{s}_{xy}$ displays a complex behavior with both dramatic oscillations and sign changes, but it eventually goes to zero above 4.5 eV.  For $p$-doped monolayer MoS$_2$, $\sigma^{s}_{xy}$ has two large peaks, respectively, with positive value of 0.29 $e^2/\hbar$ at $-1.52$ eV and negative value of $-0.32$ $e^2/\hbar$ at $-2.64$ eV.  In order to analyze the cause of the large peaks, we take $P_1$ and $P_2$ as examples, indicated in Fig.~\ref{fig:shc}(a).  When the Fermi level is located at the positions of $P_1$ and $P_2$, the band structures and spin Berry curvatures are shown in Fig.~\ref{fig:shc}(b) and \ref{fig:shc}(c), respectively.  We can see that $\Omega^s(\bm k)$ is often peaked at the places where the Fermi level crosses some tiny band-gaps induced by the SOC.  Similar behavior of $\Omega^s(\bm k)$ for the other large peaks of the $\sigma^{s}_{xy}$ are also found.  The appearance of large peaks of the positive (negative) $\Omega^s(\bm k)$ leads to the positive (negative) peaks of the $\sigma^{s}_{xy}$.

Although giant $\sigma^{s}_{xy}$ can be realized at those peak positions, such a high level of doping is unrealistic in experimental situations.  As indicated in Fig.~\ref{fig:shc}(d), the Fermi level position of $P_1$ and $P_2$ are 2.19 eV and $-1.52$ eV, respectively, corresponding to electron concentration $n_e$=1.73 $e/$cell ($\cong$ 1.96$\times$10$^{15}$ cm$^{-2}$) and hole concentration $n_h$=3.18 $e/$cell ($\cong$ 3.60$\times$10$^{15}$ cm$^{-2}$).  This could be difficult in experimental conditions either by chemical adsorption or by gate voltage.  For example, the highest carrier concentration in two-dimensional graphene is only up to 10$^{13}$ cm$^{-2}$~(Ref.~\onlinecite{Novoselov2005,Novoselov2004}).  In contrast, the low doping regimes just above the CBM and below the VBM are more relevant in experiments, as indicated by red circles in Fig~\ref{fig:shc}(a) and ~\ref{fig:shc}(d).  In the following, we only focus on this regime.

Figure~\ref{fig:shc-2}(a)-(d) show the $\sigma^{s}_{xy}$ in the low doping regimes for monolayer MoS$_2$, MoSe$_2$, WS$_2$, and WSe$_2$, respectively.  We can clearly see that $\sigma^{s}_{xy}$ increases with the SOC strength as the atoms becomes heavier, and the $p$-doped samples generally has larger $\sigma^{s}_{xy}$ than the $n$-doped samples.  This is due to the large spin splitting at the VBM [see Fig.~\ref{fig:band}(b)].  Focusing on the valence bands, we find $\sigma^{s}_{xy}$ is proportional to the Fermi energy.  The extracted slopes are listed in Table~\ref{tab:slope} together with the results from the $k\cdot p$ model.~\cite{Xiao2012}.  When the Fermi level lies inside the spin splitting gaps, the hole concentration is on the order of 10$^{13}$ cm$^{-2}$, which is realistic for experiments.  Taken $n_{h}$=1.0$\times$10$^{13}$ cm$^{-2}$ as an example, we list the calculated $\sigma^{s}_{xy}$ from both present first-principles calculation and the $k\cdot p$ model~\cite{Xiao2012} in Table~\ref{tab:slope}.  The ISHC $\sigma^{s}_{xy}$ listed in Table~\ref{tab:slope} are also comparable to those in $p$($n$)-doped semiconductors GaAs, Si, Ge, and AlAs.~\cite{Yao2005,Guo2005}

\begin{figure}
\includegraphics[width=\columnwidth]{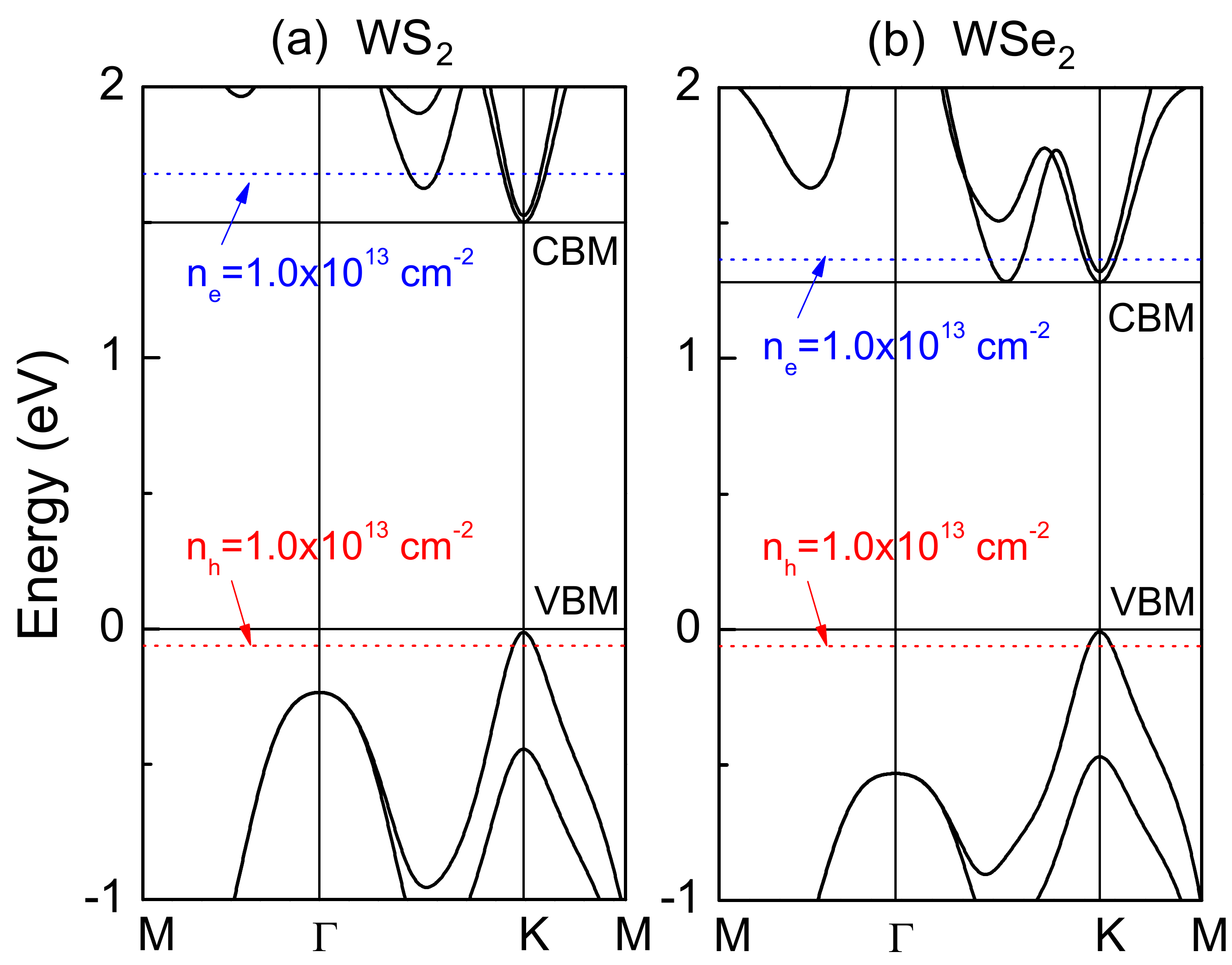}
\caption{(Color online)  The band structure of WS$_2$ and WSe$_2$ monolayers.  The CBM of WSe$_2$ monolayer still locates at $K$ point.  The dashed line indicate the position of the Fermi level at carrier concentration of $1.0\times 10^{13}$ cm$^{-2}$ for both $p$- and $n$-doped samples.}
\label{fig6}
\end{figure}

Two remarks are in order.  First, here we only compared the ISHC for $p$-doped samples for both first-principles calculations and the $k\cdot p$ method.  For $n$-doped samples, the situation is more complicated.  As we can see in Fig.~\ref{fig6}, in both WS$_2$ and WSe$_2$ the conduction band has a second local minimum between $\Gamma$ and $K$, which is very close to the band-edge at $K$.  Even under light doping ($\sim 1.0\times 10^{13}$ cm$^{-2}$), both minima will be occupied and contribute to the total ISHC, rendering the comparison between first-principles and $k\cdot p$ method meaningless.  This shows the limitation of the $k\cdot p$ method: it only captures the physics around the $K$ point, and first-principles study give us a more complete picture.  We also note that there is a small spin splitting at the CBM for both WS$_2$ and WSe$_2$.  This is due to the much heavier W atom compared to Mo (SOC scales as $Z^4$, where $Z$ is the atomic number).  In the $k\cdot p$ model, this splitting can be taken into account by considering the second-order effect due to the SOC.  Second, one may notice that $\sigma^{s}_{xy}$ is nonzero in the band-gap.  These nonzero values are not due to numerical errors, but actually reflect the finite hybridizations in real materials, similar to what has been reported in GaAs and Si.~\cite{Yao2005}  The $MX_{2}$ monolayer studied here can be viewed as a generalization of the concept of spin Hall insulator proposed by Murakami \textit{et al.},~\cite{Murakami2004} such as PbTe, which is a conventional band insulator but has nonzero $\sigma^{s}_{xy}$ without any doping.

As a comparison we also calculate $\sigma^{s}_{xy}$ for bulk MoS$_2$, MoSe$_2$, WS$_2$, and WSe$_2$, shown in Fig.~\ref{fig:shc-2}(e)-(h).  The shapes of $\sigma^{s}_{xy}$ for all bulk dichalcogenides look rather similar, though the magnitudes and positions of the peaks may differ.  In bulk, $\sigma^{s}_{xy}$ is zero immediately below the VBM, which is different from the immediate increasing of the $\sigma^{s}_{xy}$ in monolayers.  The reason is that the VBM in bulk always locates at the $\Gamma$ point, which has negligibly small contribution to $\sigma^{s}_{xy}$.  To compare with monolayers, we divide the $\sigma^{s}_{xy}$ in monolayer by its thickness and find that at the same Fermi level the $\sigma^{s}_{xy}$ in bulk is about an order of magnitude smaller than that in monolayers.

Finally we mention that here we only considered the intrinsic contribution to the spin Hall effect, in which the spin Hall current is driven by the Berry curvature of the Bloch bands.  There are also extrinsic contributions coming from scattering of impurities and phonons.  When the sample is hole-doped, the effect of phonon scattering on the SHC should be weak because the phonon scattering will mostly contribute to intra-valley scattering, in which the spin $z$-component is nearly conserved due to the large spin-orbit splitting at the valence band top.  On the other hand, impurity scattering can provide the large momentum transfer needed for the inter-valley scattering, and their effect on valley-dependent transport properties remains to be investigated.

\section{Summary}

In summary, using the first-principles calculations, we have investigated the intrinsic spin Hall effect in monolayers MoS$_2$, MoSe$_2$, WS$_2$, as well as WSe$_2$, driven by valley-contrasting Berry curvature.  We find that the ISHC is comparable to that in $p$($n$)-doped semiconductors GaAs, Si, Ge, and AlAs.~\cite{Yao2005,Guo2005}  We show that the effective model may not be adequate to describe the low-energy dynamics in WS$_2$ and WSe$_2$ monolayers. We also calculated the ISHC in inversion-symmetric bulk systems and find that it is about an order of magnitude smaller than the ISHC in monolayers.  The large ISHC, plus other interesting physical properties of these materials, such as giant spin splitting at VBM~\cite{Zhu2011} and valley-selective circular dichroism,~\cite{Xiao2012,Zeng2012,Mak2012,Cao2012} characterize these materials as an exciting platform for the application of the valleytronics and spintronics.

\begin{acknowledgments}
W.F. was partly supported by the Laboratory Directed Research and Development Program of ORNL.  W.Y. was supported by the Research Grant Council of Hong Kong under Grants No. HKU706412P, Y.Y. by the MOST Project of China (Grants No. 2011CBA00100) and NSF of China (Grants No. 10974231 and 11174337), and W.Z. and D.X. by the U.S. Department of Energy, Office of Basic Energy Sciences, Materials Sciences and Engineering Division.  We also thank Supercomputing Center of Chinese Academy of Sciences (SCCAS) and Texas Advanced Computing Center (TACC) for the computational supports.
\end{acknowledgments}


\begin{thebibliography}{100}
\bibitem{foot} The spin Hall effect discussed here is driven by the valley-contrasting Berry curvature and requires inversion symmetry breaking.  In general the spin Hall effect does not require inversion symmetry breaking.
\end{thebibliography}
\end{document}